\documentclass[aps,prl,floatfix,twocolumn,superscriptaddress,preprintnumbers,nofootinbib]{revtex4-1}

\usepackage{textcomp} 
\usepackage{slashed}
\usepackage{epsfig,latexsym,cancel,amssymb,amsmath,verbatim,mathrsfs}
\usepackage{color}
\usepackage{graphicx}
\usepackage{enumitem}
\usepackage[colorlinks,citecolor=blue]{hyperref}

\newcommand{\beq}{\begin{equation}}
\newcommand{\eeq}{\end{equation}}
\newcommand{\bea}{\begin{eqnarray}}
\newcommand{\eea}{\end{eqnarray}}

\newcommand{\nn}{\nonumber}
\newcommand{\tttt}{t\bar{t}t\bar{t}}
\newcommand{\htt}{Htt}

\allowdisplaybreaks[4]

\begin{document}

\title{Limiting Top-Higgs Interaction and Higgs-Boson Width from Multi-Top Productions}

\author{Qing-Hong Cao}
\email{qinghongcao@pku.edu.cn}
\affiliation{Department of Physics and State Key Laboratory of Nuclear Physics and Technology, Peking University, Beijing 100871, China}
\affiliation{Collaborative Innovation Center of Quantum Matter, Beijing 100871, China}
\affiliation{Center for High Energy Physics, Peking University, Beijing 100871, China}

\author{Shao-Long Chen}
\email{chensl@mail.ccnu.edu.cn}
\affiliation{Key Laboratory of Quark and Lepton Physics (MoE) and Institute of Particle Physics, Central China Normal University, Wuhan 430079, China}
\affiliation{Center for High Energy Physics, Peking University, Beijing 100871, China}

\author{Yandong Liu}
\email{ydliu@bnu.edu.cn}
\affiliation{Key Laboratory of Beam Technology of Ministry of Education, College of Nuclear Science and Technology, Beijing Normal University, Beijing 100875, China}
\affiliation{Beijing Radiation Center, Beijing 100875, China}

\author{Rui Zhang}
\email{zhang.rui@pku.edu.cn}
\affiliation{Department of Physics and State Key Laboratory of Nuclear Physics and Technology, Peking University, Beijing 100871, China}

\author{Ya Zhang}
\email{zhangya1221@pku.edu.cn}
\affiliation{Department of Physics and State Key Laboratory of Nuclear Physics and Technology, Peking University, Beijing 100871, China}

\begin{abstract}
We demonstrate that the multi-top productions efficiently probe the CP-property of top-Higgs interaction and the Higgs-boson width at the LHC. The four top-quark production alone can exclude a purely CP-odd top-quark Yukawa coupling at the 13~TeV LHC with an integrated luminosity of $430~{\rm fb}^{-1}$, regardless the size of the Yukawa coupling. Combining the single Higgs-boson production, the $t\bar{t}H$ associated production and the four top-quark production, we show that the CP-phase of the top-quark Yukawa coupling and the Higgs-boson width can be stringently bounded at the LHC with integrated luminosities of $300~{\rm fb}^{-1}$ and $3000~{\rm fb}^{-1}$. 

\end{abstract}

\maketitle

\noindent\textbf{1. Introduction.}

Measurement of the rate of the association production of a top-quark pair and a Higgs boson (named as the $\htt$ channel) provides a direct test of the Higgs-boson interaction with the top-quark, i.e. the so-called top-quark Yukawa interaction. Recently, both the ATLAS and CMS collaborations confirm the top-quark Yukawa coupling ($y_t$) at the $5\sigma$ C.L.~\cite{Aaboud:2018urx,Sirunyan:2018hoz,Sirunyan:2018koj}. The Next task is to test the CP property of $y_t$~\cite{Kobakhidze:2014gqa}. As the strength of $y_t$ is comparable to the coupling strength of strong interaction ($g_s$), it has been shown that four top-quark ($\tttt$) production can probe the CP-even $y_t$ coupling and also the tiny Higgs-boson width at the LHC and future colliders~\cite{Cao:2016wib,Sirunyan:2017roi,Frederix:2017wme,Contino:2016spe}. In the work we first illustrate that the $\tttt$ production is good at probing the CP-property of the $y_t$ coupling, and then perform a global fit to explore the sensitivity of the 13~TeV LHC on $y_t$ after combining the single-Higgs production and $\htt$ production. 

The Higgs-boson width plays a crucial role in all the Higgs-boson productions. Even though very tiny ($\sim 4~{\rm MeV}$) and hard to be measured, the Higgs-boson width could distinguish various new physics models, e.g. the composite Higgs model or  the invisible decays of the Higgs-boson in the Higgs-portal dark matter models, etc. As the Higgs width is much less than the detector energy resolution $\sim1~{\rm GeV}$, one can extract it through the comparison of the processes involving the on-shell and off-shell Higgs boson. One way is to examine the on-shell and off-shell Higgs-boson effect in the $gg\to H\to ZZ^*$ process~\cite{Caola:2013yja}. Another way is to compare the $t\bar{t}H$ and $t\bar{t}t\bar{t}$ production~\cite{Cao:2016wib}. In this study we explore the sensitivity of the LHC to the Higgs-boson width through the multi-top productions.

\begin{figure}
\includegraphics[scale=0.33]{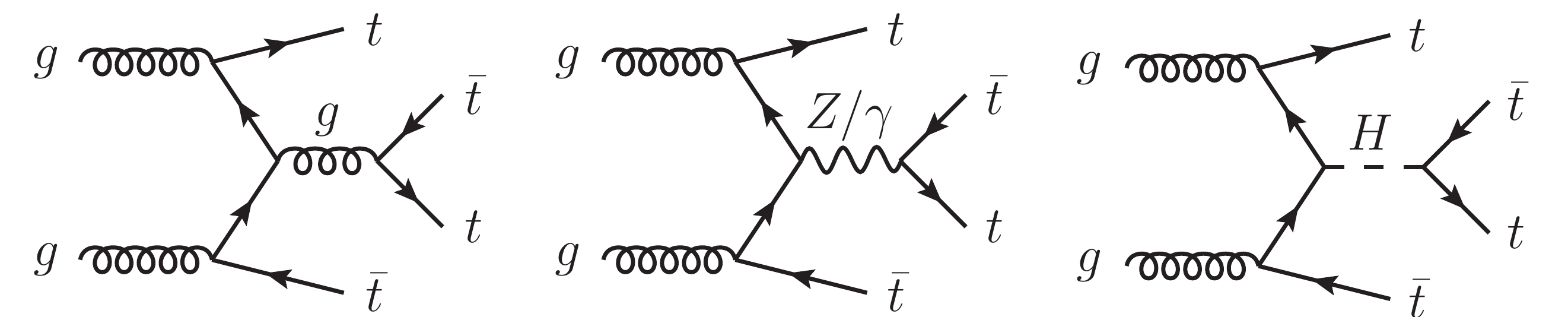}
\caption{Illustrative Feynman diagrams of $\tttt$ productions.}
\label{fig:feyn}
\end{figure}

The general top-quark Yukawa coupling $y_t$ is parameterized as following
\beq
\mathcal{L}_{\htt} = -\frac{m_t}{v}H\bar{t} (a_t+ib_t\gamma_5) t~,
\eeq
where $m_t$ is the top-quark mass, $v(=174~{\rm GeV})$ the vacuum expectation value, and the coefficient $a$ ($b$) denotes the CP-even (CP-odd) coupling, respectively. Figure~\ref{fig:feyn} displays The representative Feynman diagrams of the $\tttt$ production, which occurs either through the gluon mediation, the electroweak gauge-boson mediation, or the Higgs boson mediation in the SM. We name the corresponding matrix elements as $\mathcal{M}_{g}$, $\mathcal{M}_{Z/\gamma}$, and $\mathcal{M}_{H}$.
The contribution of $\mathcal{M}_H$ to the $\tttt$ production cross section is proportional to $y_t^4$, and that enables us to measure $y_t$ through the $\tttt$ channel. Since the QCD and electroweak gauge interactions of top quarks have been well established, we consider only the top Yukawa coupling might differ from the SM value. The cross-section of $\tttt$ production is parameterized as follows: 
\beq
\sigma(\tttt) = \sigma(\tttt)_{g+Z/\gamma} + \sigma(\tttt)_{\rm int} +  \sigma(\tttt)_H,
\eeq
where
\bea
\sigma(\tttt)_{g+Z/\gamma}&~\propto~& \left|\mathcal{M}_{g} + \mathcal{M}_{Z/\gamma}\right|^2, \nn\\
\sigma(\tttt)_{\rm int} &\propto& \mathcal{M}_{g+Z/\gamma}\mathcal{M}^\dagger_{H}+\mathcal{M}^\dagger_{g+Z/\gamma}\mathcal{M}_{H},\nn\\
\sigma(\tttt)_{H}&\propto& \left|\mathcal{M}_{H}\right|^2.
\eea
We calculate $\sigma(\tttt)$ at the leading order using MadEvent~\cite{Alwall:2007st}, which yields
\bea
\sigma(\tttt)_\text{13~TeV}&=&9.998-1.522a_t^2+2.883b_t^2 \nn\\
&+& 1.173a_t^4 + 2.713a_t^2b_t^2+1.827b_t^4,
\eea
where we ignore mixing terms of $a_tb_t$, $a_t^3b_t$ and $a_tb_t^3$ as their contributions are about one thousandth of those terms shown above.   
The first term on the right-handed side denotes $\sigma(\tttt)_{g+Z/\gamma}$, the second and third terms represent $\sigma(\tttt)_{\rm int}$, and the last three terms label $\sigma(\tttt)_{H}$.

It is worth noting that, in the interference contribution $\sigma^{\rm SM}(\tttt)_{\rm int}$, the CP-even contribution ($\propto a_t^2$) is negative while the CP-odd contribution ($\propto b_t^2$) is positive. In the $\sigma^{\rm SM}(\tttt)_{H}$, the CP-even and CP-odd contributions as well as their interference are all positive. Moreover, the CP-odd contribute is larger than the CP-even contrition in the $\sigma^{\rm SM}(\tttt)_{\rm int}$ and $\sigma^{\rm SM}(\tttt)_{H}$; therefore, it is more promising to observe the pseudo-scalar effect through the $\tttt$ production. 

The QCD corrections to the signal process are large and yield a $k$-factor ($K_F$) as 1.58 for the SM ($a_t=1$, $b_t=0$)~\cite{Frederix:2017wme}. In the study we rescale the LO cross section of the signal process by an overall $K_F=1.58$ to mimic the high order QCD effects in both the CP-even and CP-odd cases. The drawback of the $\tttt$ production is that it suffers from a large dependence on the choices of the renormalization scale and the factorization scale~\cite{Bevilacqua:2012em,Alwall:2014hca}. Even worse, including higher order quantum corrections only improves the scale dependence mildly~\cite{Frederix:2017wme}. We conservatively consider the cross section of the signal process exhibits a $50\%$ uncertainty of the scale dependence throughout our study, i.e. \beq
\sigma=\sigma_0 \pm \delta\sigma_\mu,\qquad \delta\sigma_\mu = 50\% \times \sigma_0,
\eeq 
where $\sigma_0$ denotes the central value of the cross section and $\delta\sigma_\mu$ is the variation from the scale dependence.  

\noindent\textbf{2. Purely CP-odd $y_t$ coupling.}

In the previous study~\cite{Cao:2016wib} we explored the potential of the LHC on the CP-even $y_t$, considering the event topology of the $\tttt$ production consists of lots of jets, multiple charged leptons and missing transverse momentum. Such a collider signature has been studied by the CMS collaboration~\cite{Sirunyan:2017roi} to constrain a purely CP-even $y_t$. Now we first extend our study to the purely CP-odd $y_t$ and then address on the CP-mixture case.  

\begin{figure}
\includegraphics[scale=0.6]{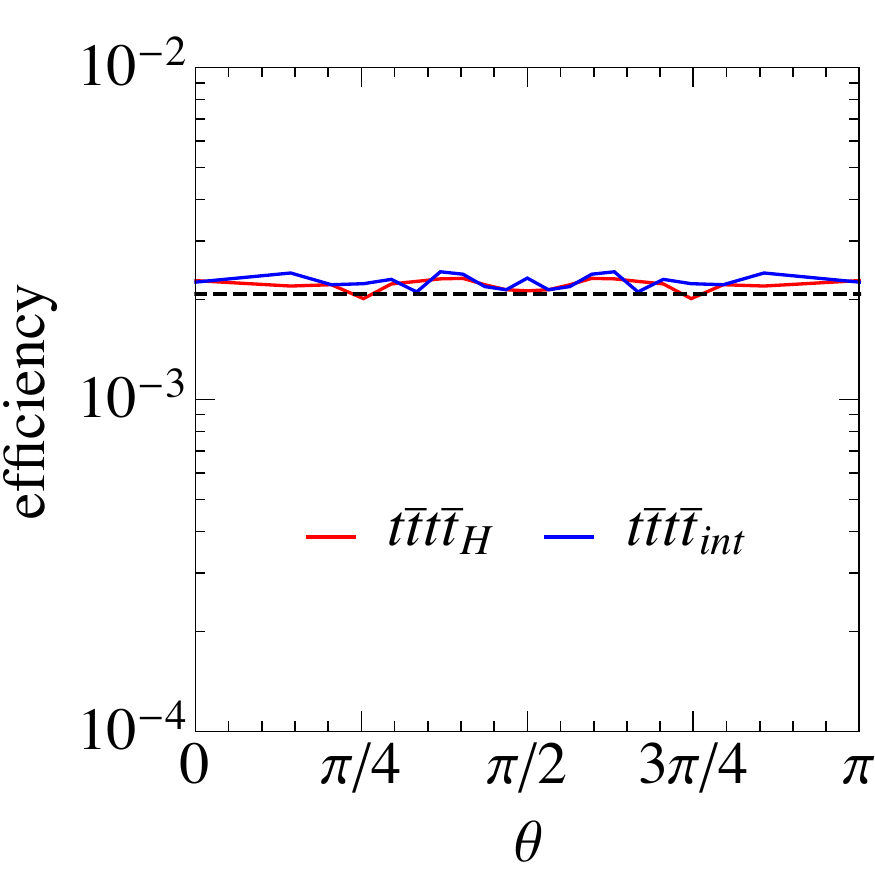}
\caption{The cut efficiency of the interference contribution (blue) and the Higgs-mediation contribution alone (red) in the $t\bar{t}t\bar{t}$ production as a function of the CP-phase $\theta$.}
\label{fig:cut_eff}
\end{figure}

The kinematics distributions of the final state particles are affected only mildly by the CP-phase such that the top-Higgs couplings with different CP-phases yield almost the same cut efficiencies. Figure~\ref{fig:cut_eff} displays the cut efficiencies of the $\tttt$ production as a function of the CP-phase $\theta$ of $y_t$, where 
\beq
\theta\equiv \arctan (b_t/a_t).
\eeq
The horizontal dashed curve denotes the cut efficiency of gauge boson mediation contribution.
While, the blue and red curve represents the cut efficiency of the interference contribution and the Higgs-boson mediation contribution, respectively. The three curves are close together in the whole range of CP-phase $\theta$.  
Therefore we can use the same strategy from the experimental collaboration to probe the potential of the LHC on the $y_t$.
In our study we adopt the cut flow used by the CMS collaboration~\cite{Sirunyan:2017roi}, which is optimized for searching the SM $\tttt$ production. The CMS study shows that there are 0.7 events of the SM background after all the cuts at the 13~TeV LHC with an integrated luminosity ($\mathcal{L}$) of $35.9~{\rm fb}^{-1}$~\cite{Sirunyan:2017roi}. We thus estimate the number of  the SM background events ($N_b$) as $N_b=0.0195\mathcal{L}$ at the 13~TeV LHC.

\begin{figure}
\includegraphics[scale=0.4]{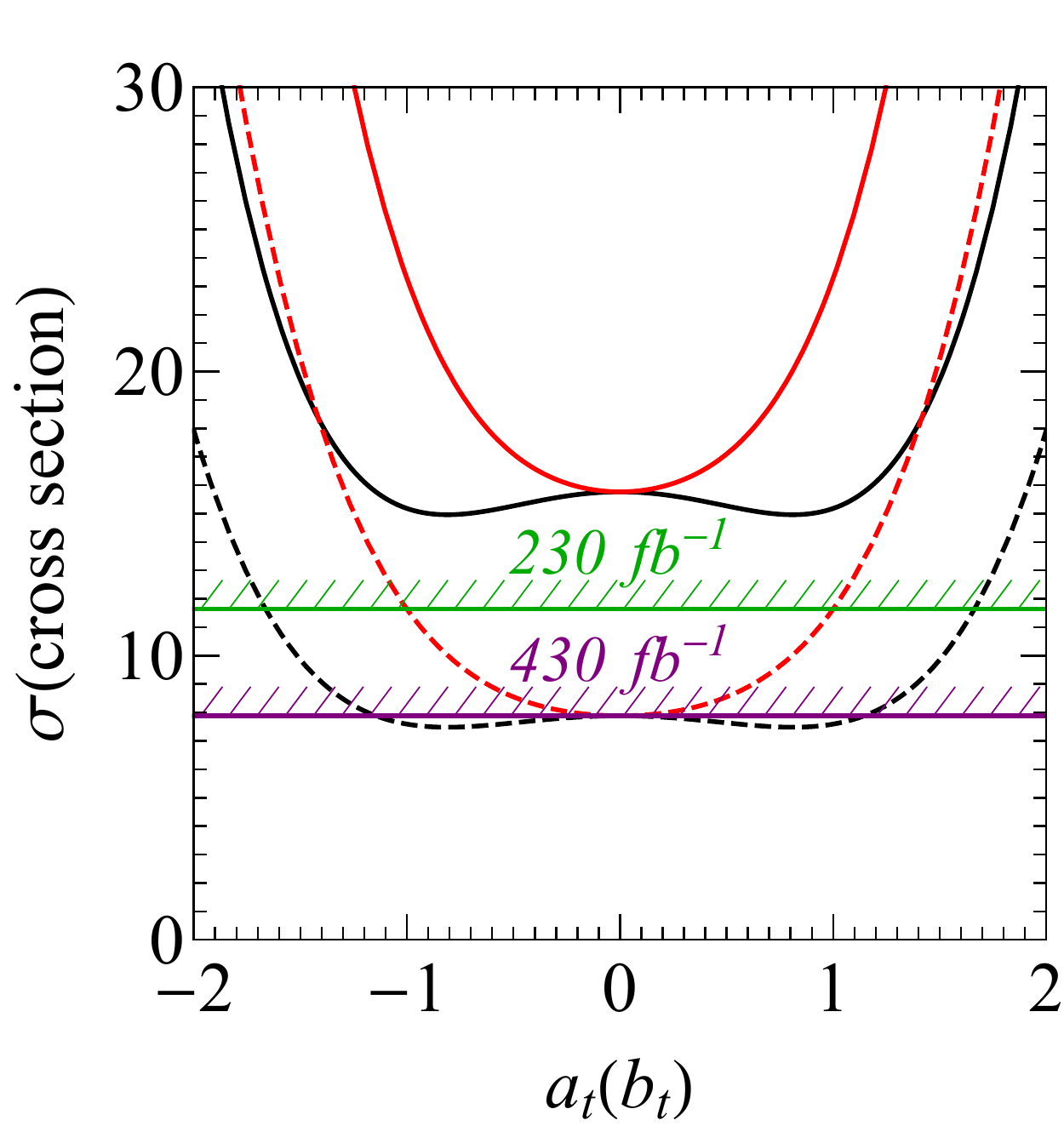}
\caption{Cross section of the $\tttt$ production as a function of a purely CP-even coupling $a_t$ (black) or a purely CP-odd coupling $b_t$ (red). The solid curves represent $\sigma_0$ while the dashed curves denote $\sigma_0-\delta\sigma_\mu$. The horizontal lines represent the upper limit of the production rate with an integrated luminosity of $230~{\rm fb}^{-1}$ (green) and $430~{\rm fb}^{-1}$ (purple). }
\label{fig:ex0}
\end{figure}

\begin{figure*}
\includegraphics[scale=0.5]{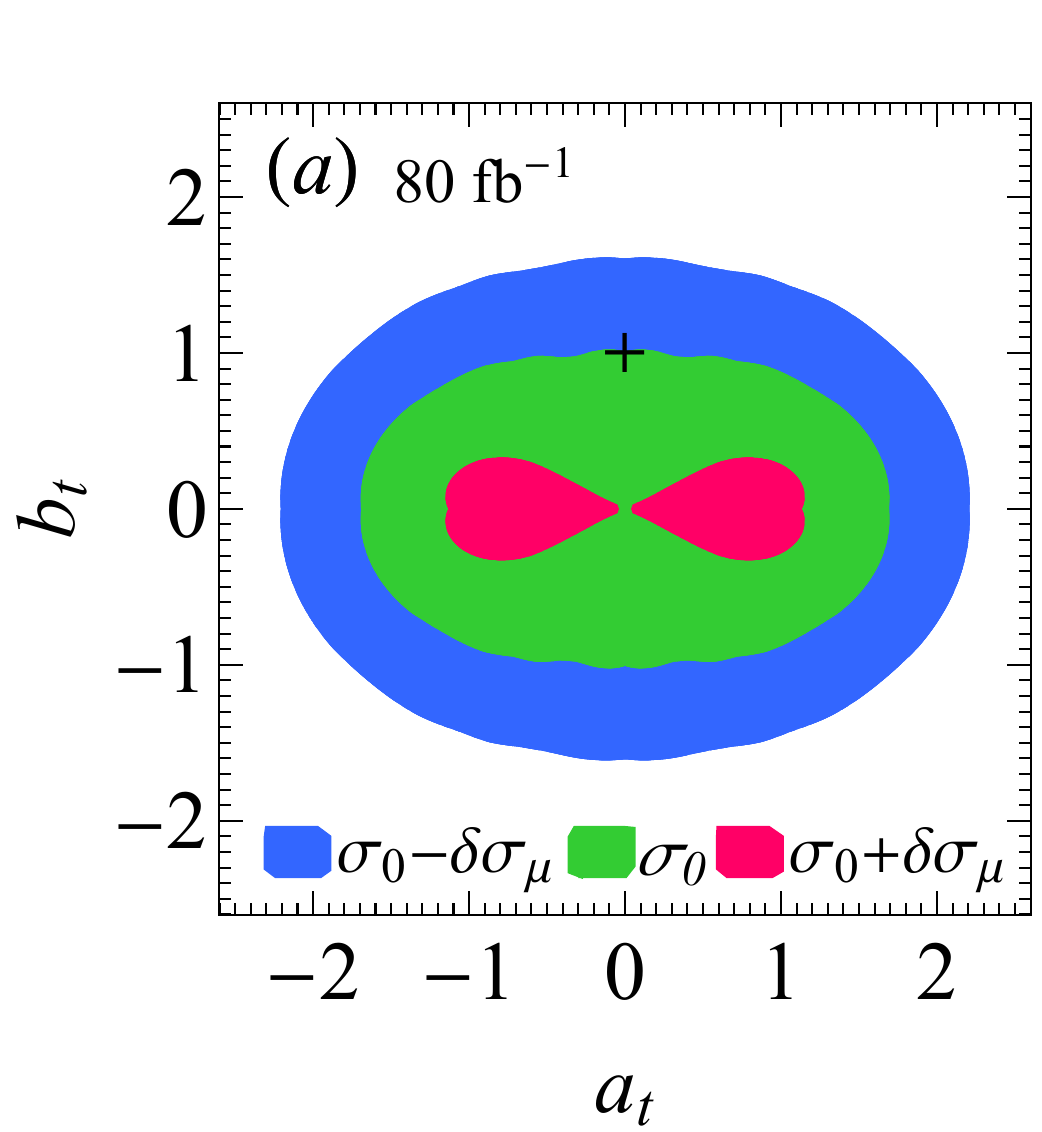}
\includegraphics[scale=0.5]{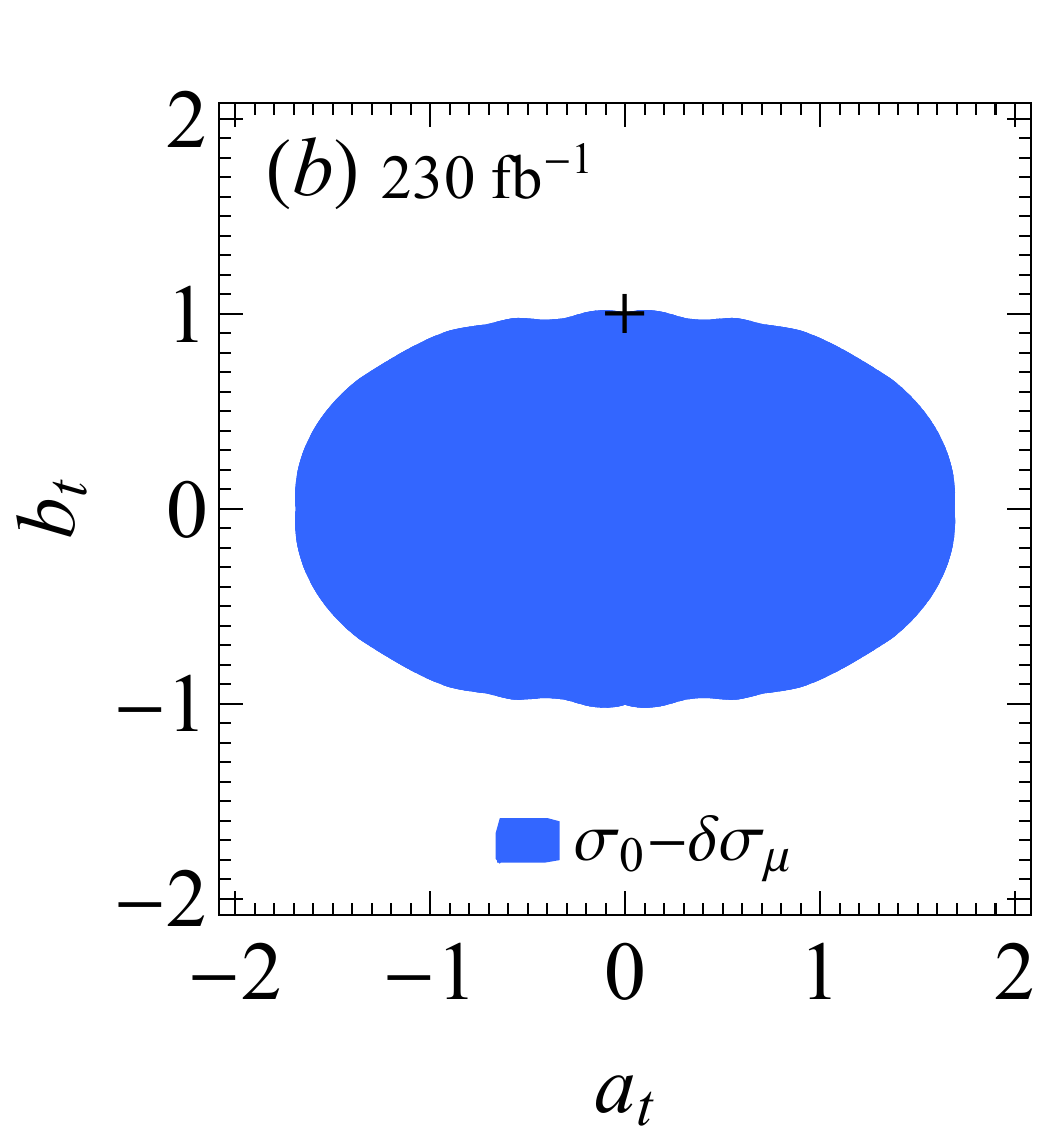}
\includegraphics[scale=0.5]{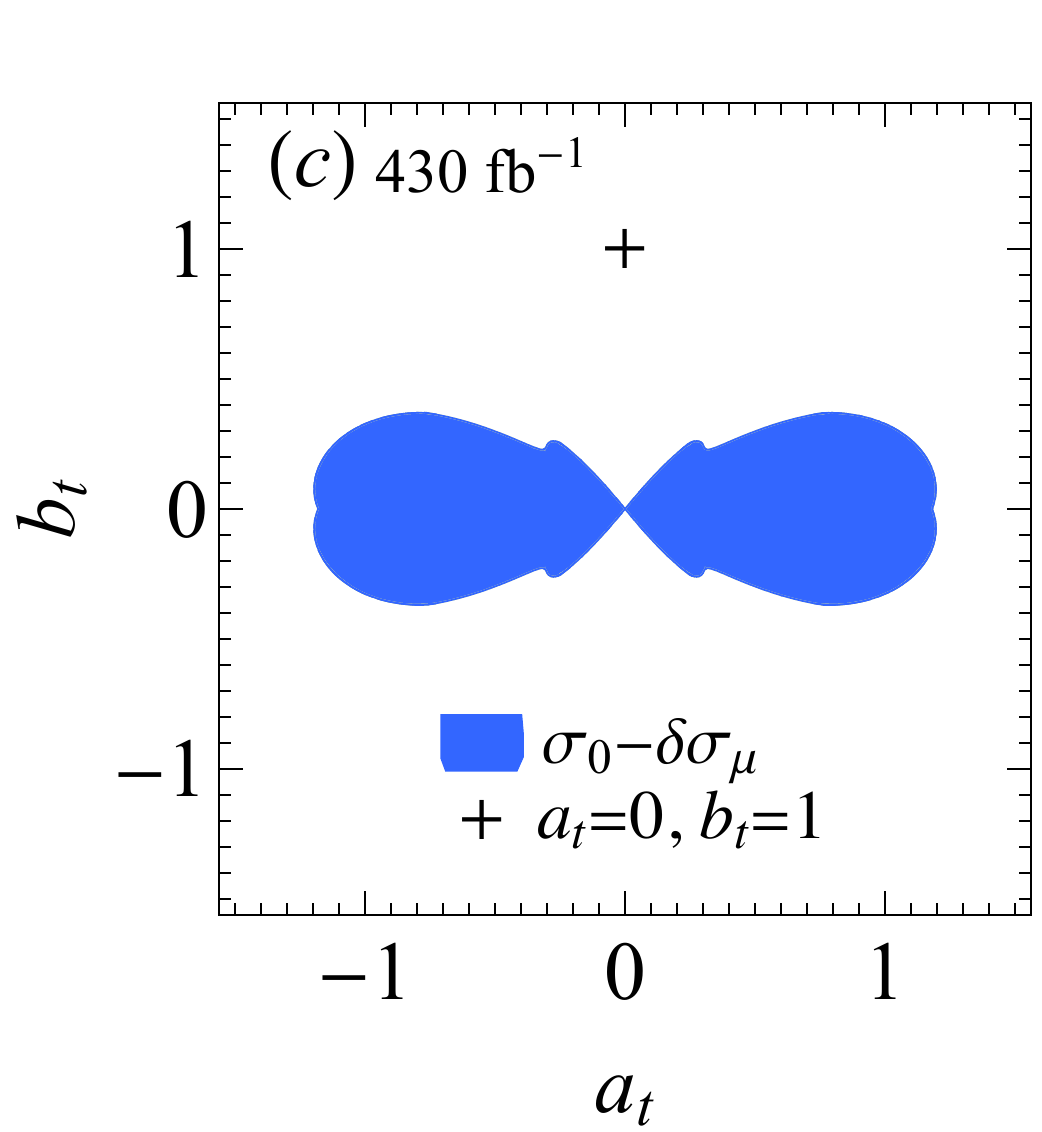}
\caption{Allowed parameter space in the plane of $a_t$ and $b_b$ at the $95\%$ confidence level with an integrated luminosity of $80~{\rm fb}^{-1}$ (a), $230~{\rm fb}^{-1}$ (b) and $430~{\rm fb}^{-1}$ (c) if a null result of the $\tttt$ production is reported at the 13 TeV LHC. The symbol ``$+$" denotes the purely CP-odd coupling ($a_t=0$, $b_t=1$). }
\label{fig:ex}
\end{figure*}

Next we simulate the signal process of a purely CP-odd $y_t$ coupling ($a_t=0$ and $b_t=1$) following the CMS strategy. 
Our simulation shows that 3.97 events of the signal ($N_s$) and 1.56 events of the background survive all the cuts with the current integrated luminosity of 80 fb$^{-1}$. 
As there are few events of both the signal and the backgrounds after the kinematics cuts, we estimate the significance of excluding the purely CP-odd $y_t$ coupling using~\cite{Cowan:2010js}
\beq \label{statistic:ex}
\sqrt{-2\left[N_b \log \frac{N_s+N_b}{N_b} -N_s\right]}=2.
\eeq
Figure~\ref{fig:ex0} displays the production cross section of the signal process as a function of the $y_t$ coupling, where the red and black curves denote the purely CP-odd coupling $a_t$ and the purely CP-even coupling $b_t$, respectively. As mentioned above, the $t\bar{t}t\bar{t}$ production suffers from large scale dependences. For illustration we plot both the central value of the signal production rate $\sigma_0$ (solid-curve) and the lower value $\sigma_0-\delta\sigma_\mu$ (dashed-curve). Those horizontal lines show the projected upper limit of the cross section of the $t\bar{t}t\bar{t}$ production obtained at the 13~TeV LHC with an integrated luminosity of $230~{\rm fb}^{-1}$ (green) and $430~{\rm fb}^{-1}$ if only null result is found in the four-top measurement.

Figure~\ref{fig:ex} displays the allowed parameter space in the plane of $a_t$ and $b_t$ at the 95\% confidence level for various integrated luminosities: (a) $80~{\rm fb}^{-1}$, (b) $230~{\rm fb}^{-1}$ and (c) $430~{\rm fb}^{-1}$. The cross symbol denotes the purely CP-odd coupling ($a_t=0, b_t=1$). The shaded regions are still allowed if the $t\bar{t}t\bar{t}$ is not observed for a given integrated luminosity, where the blue (green, red) region is obtained when considering the lower, central and upper value of the signal production rate, respectively. 
Even though the current data cannot probe the purely CP-odd coupling when taking the scale uncertainty of the signal production into account, as shown in Fig.~\ref{fig:ex}(a), collecting more data is able to constrain the CP-odd coupling. For example, one can exclude the purely CP-odd coupling ($a_t=0,b_t=1$) at the 2$\sigma$ C.L.,  even considering the lower value of the signal production, when an integrated luminosity of 230 fb$^{-1}$ is accumulated; see Fig.~\ref{fig:ex}(b). Furthermore, any pure CP-odd coupling can be excluded when the integrated luminosity reaches $432~{\rm fb}^{-1}$, which is depicted in Fig.~\ref{fig:ex}(c).

~\\
\noindent\textbf{3. Global analysis of multi-top productions.}

Now we turn to the case of CP-mixture coupling. The CP-phase of $y_t$ plays different roles in various Higgs-boson production channels, e.g. the single-Higgs production, the $t\bar{t}H$ associated production and the $t\bar{t}t\bar{t}$ production. For example, the $t\bar{t}H$ production is modified as~\cite{Boudjema:2015nda}
\beq
\frac{\sigma(gg\to t\bar{t}H)}{\sigma(gg\to t\bar{t} H)_{\rm SM}}\equiv \mu_{t\bar{t}H}^{\rm Th}= a_t^2+0.46b_t^2,
\eeq
while the single Higgs-boson production through gluon-fusion ($gg\to H$) is modified as~\cite{Boudjema:2015nda}
\beq
\frac{\sigma(gg\to H)}{\sigma(gg\to H)_{\rm SM}}\equiv \mu_{gg\to H}^{\rm Th}=a_t^2+2.29b_t^2.
\eeq
Obviously, the scalar component of the $y_t$ coupling dominates in the former process while the pseudo-scalar dominates in the latter. Therefore, combining the two Higgs-boson  production channels with the $t\bar{t}t\bar{t}$ production would yield much tighter constraints on $y_t$.  

When dealing with the signal strength measured in various Higgs-boson production channels, one needs to pay some attention to the Higgs-boson width effect, which could dramatically alter the branching ratios of the Higgs-boson decay mode. Define $R_\Gamma$ as the ratio of the Higgs-boson width ($\Gamma_H$) to the SM value ($\Gamma_H^{\rm SM}$), i.e.
\beq
R_\Gamma\equiv \Gamma_H/\Gamma_H^{\rm SM}.
\eeq
The ratio describes the new physics effects either from modification of the Higgs-boson couplings to the SM particles, or from unknown decay modes such as the invisible Higgs-boson decay in the Higgs-portal dark matter models~\cite{Patt:2006fw,Geng:1988mw}. 
Recently, the $t\bar{t}H$ production is measured in final states with electrons, muons and hadronically decaying $\tau$ leptons by the CMS collaboration~\cite{CMS:2018dmv}. As those leptonic modes are not sensitive to the top-Higgs couplings, we can parametrize the signal strength in the $t\bar{t}H$ production as 
\beq
\mu_{t\bar{t}H} = (a_t^2+0.46b_t^2)/R_\Gamma.
\label{eq:tth_th}
\eeq
On the other hand, the best mode to measure the single-Higgs production is through the di-photon mode $H\to \gamma\gamma$, which is also affected by $y_t$. Using the so-called  narrow width approximation we parametrize the signal strength of $gg\to H\to \gamma\gamma$ as 
\bea
&&\frac{\sigma(gg\to H \to \gamma\gamma)}{\sigma(gg\to H \to \gamma\gamma )_{\rm SM}}\nn\\
&=&  \left(a_t^2+2.29b_t^2\right)\times \frac{0.0237(-8.33+1.83a_t^2)+0.185b_t^2}{R_\Gamma}.\nn\\
\label{eq:ggh_th}
\eea
It is shown that a global analysis of combining multi-top productions can probe $\Gamma_H$ from above analysis.

\begin{figure}
\includegraphics[scale=0.5]{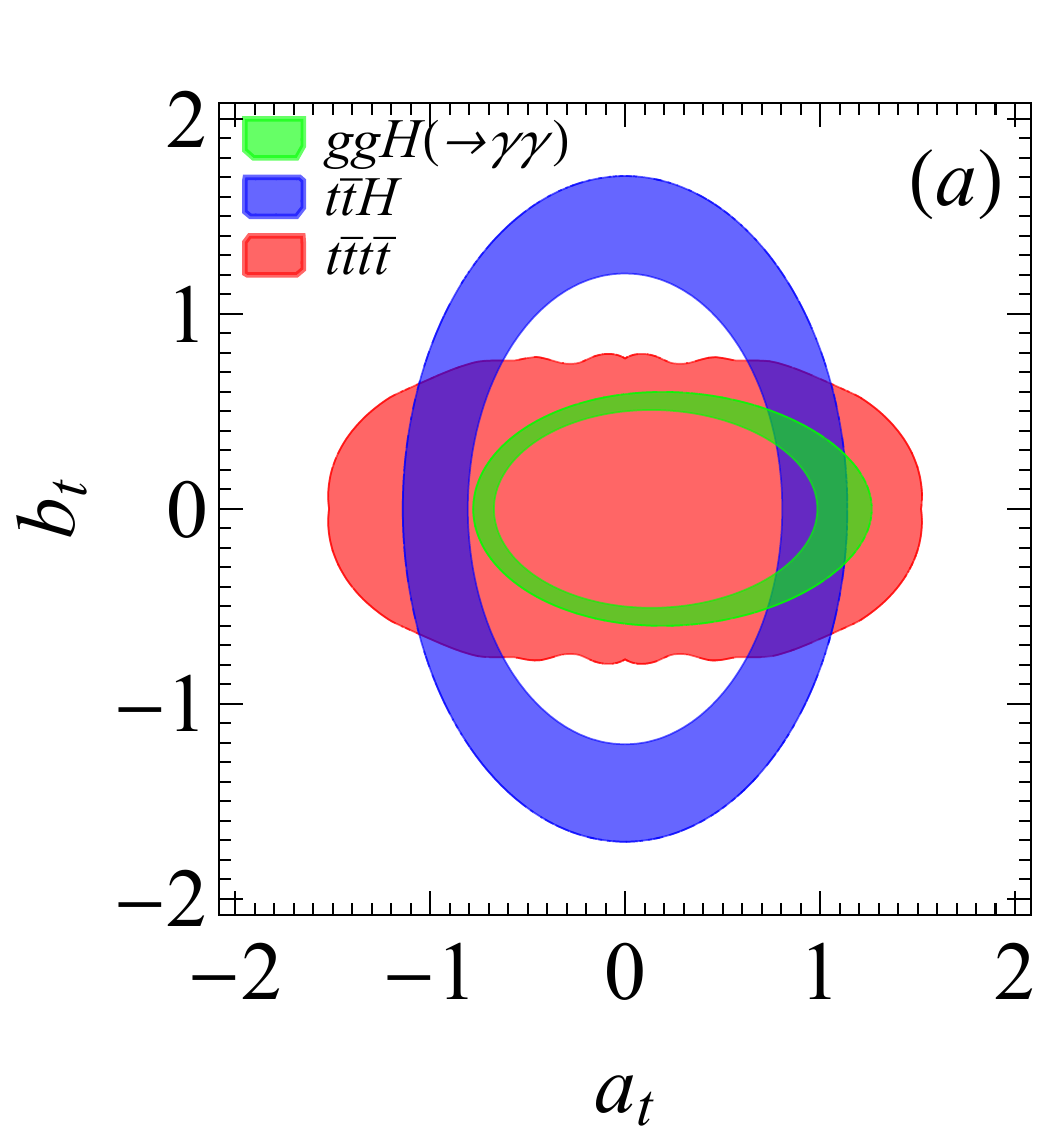}
\includegraphics[scale=0.5]{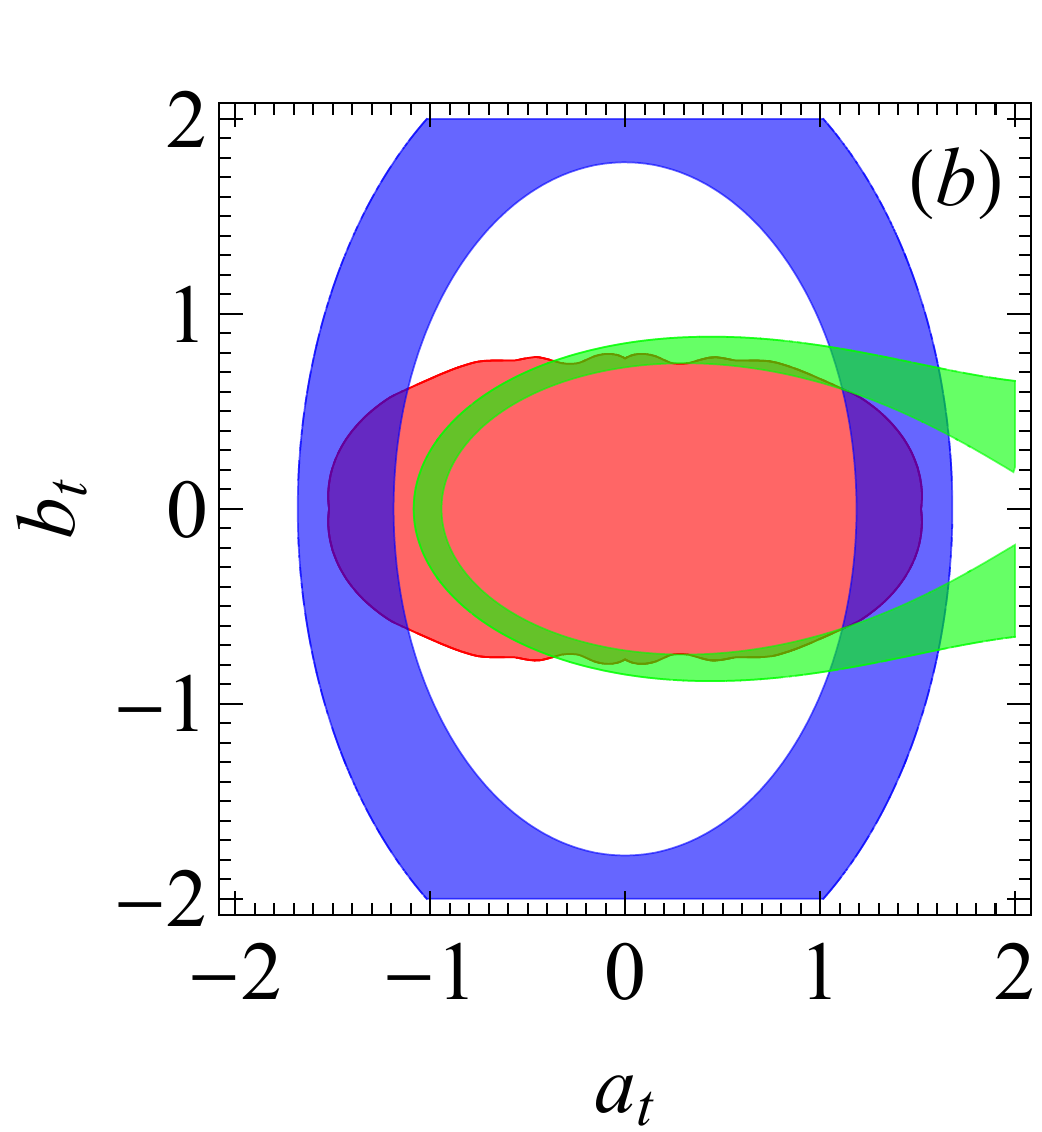}
\caption{Parameter space in the plane of $a_t$ and $b_t$ consistent with the current measurements of $ggH(\rightarrow\gamma \gamma)$ (green) and the $t\bar{t}H$ production (blue), and the projection of the $t\bar{t}t\bar{t}$ production (red) if a null result is reported at the LHC with $\mathcal{L}=300 {\rm fb}^{-1}$. We fix $R_\Gamma=1$ in the Figure (a) but relax it in the Figure (b). }
\label{fig:aliveregion:300} 
\end{figure}

\begin{figure*}
\includegraphics[scale=0.59]{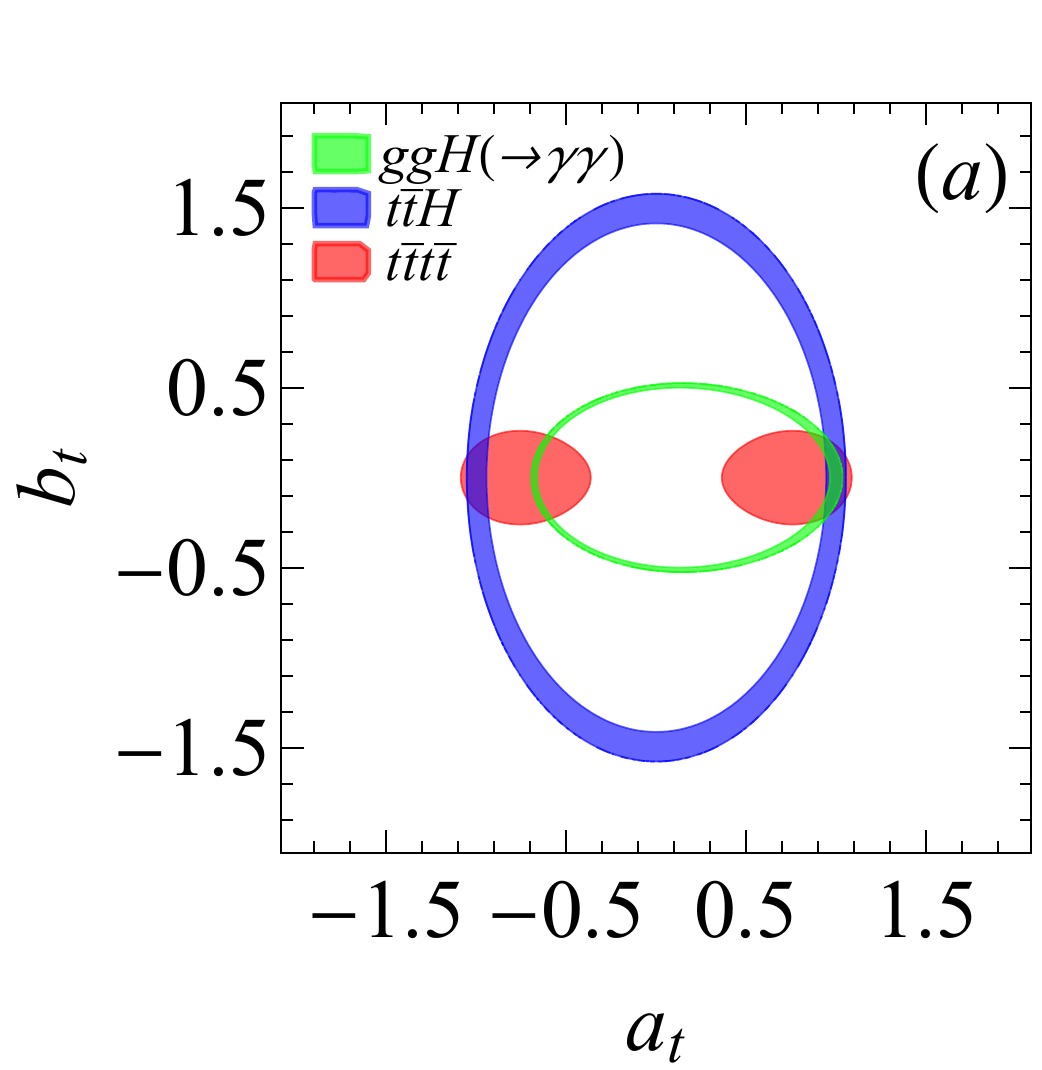}~~
\includegraphics[scale=0.6]{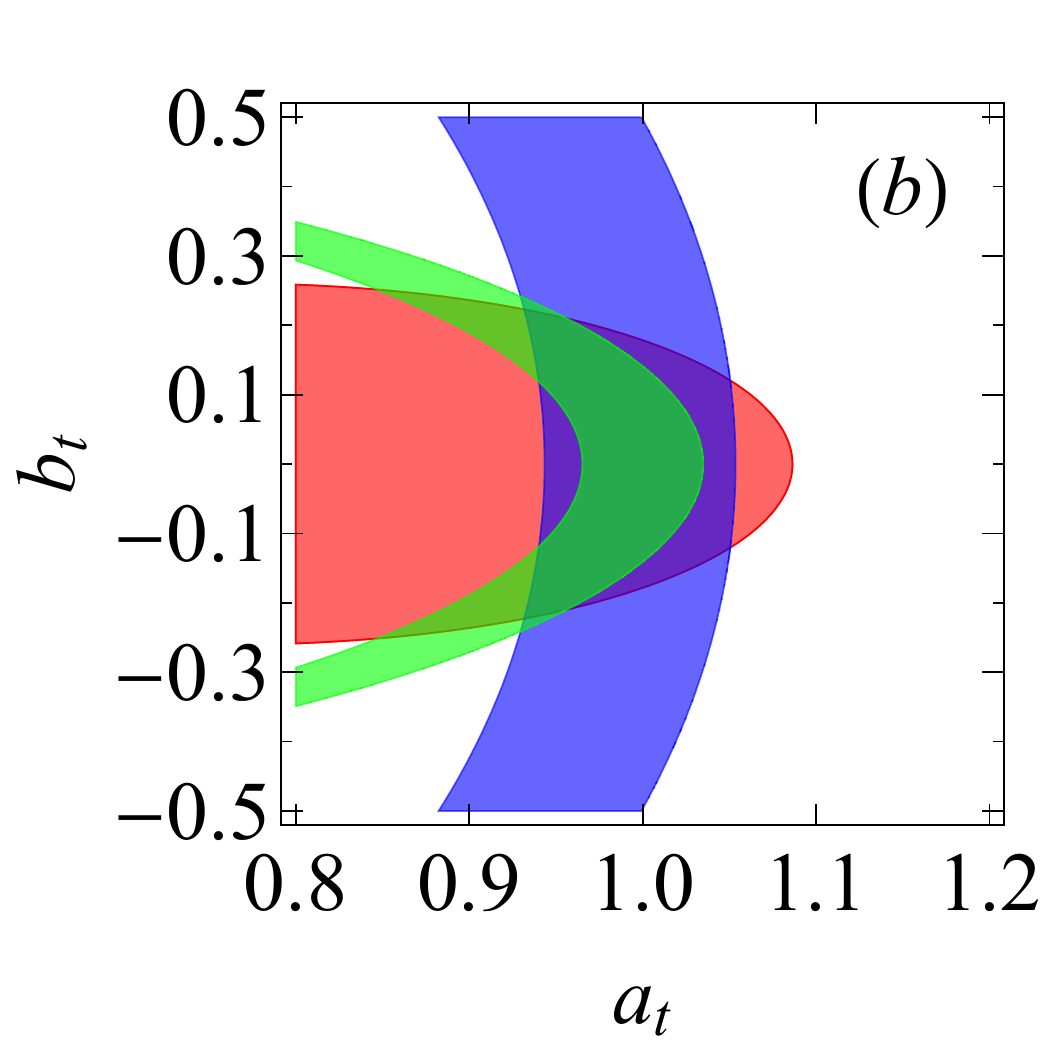}\\
\includegraphics[scale=0.6]{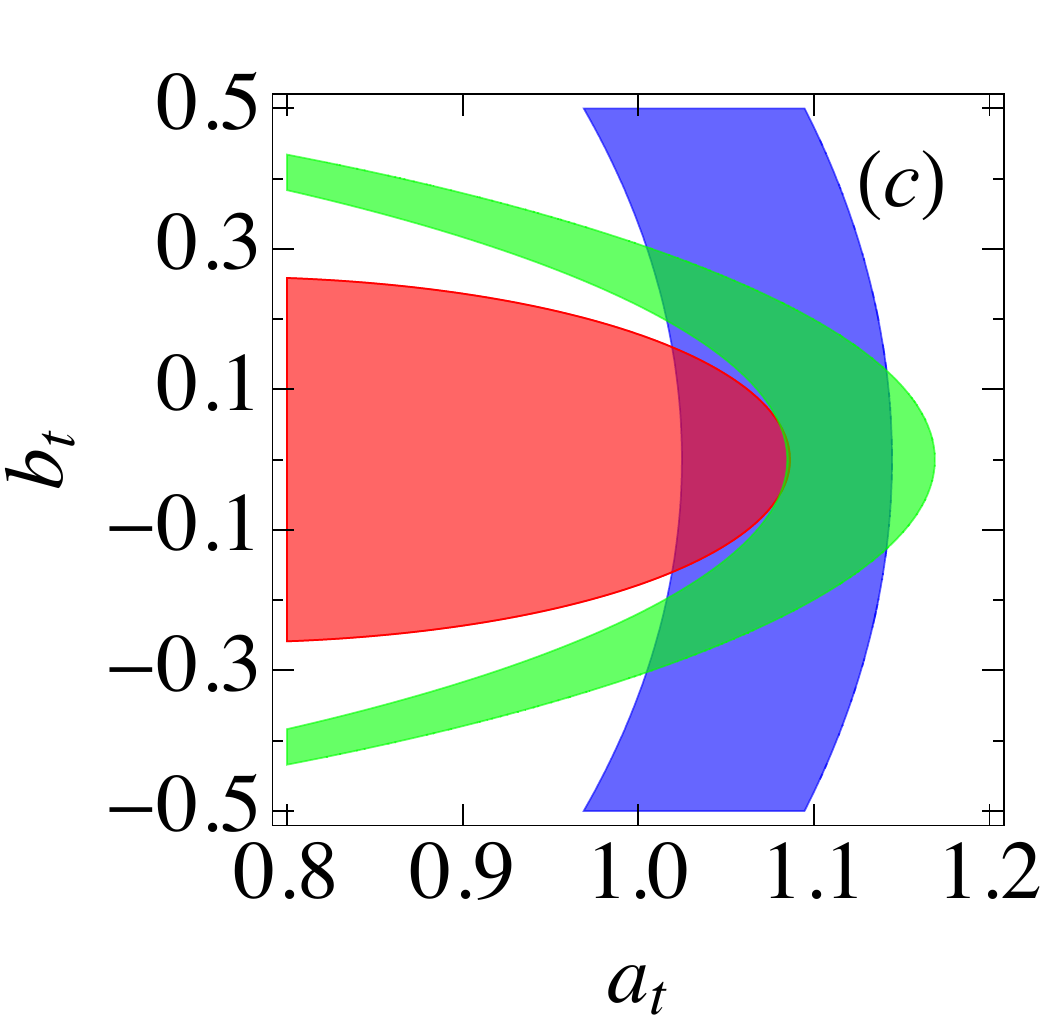}
\includegraphics[scale=0.6]{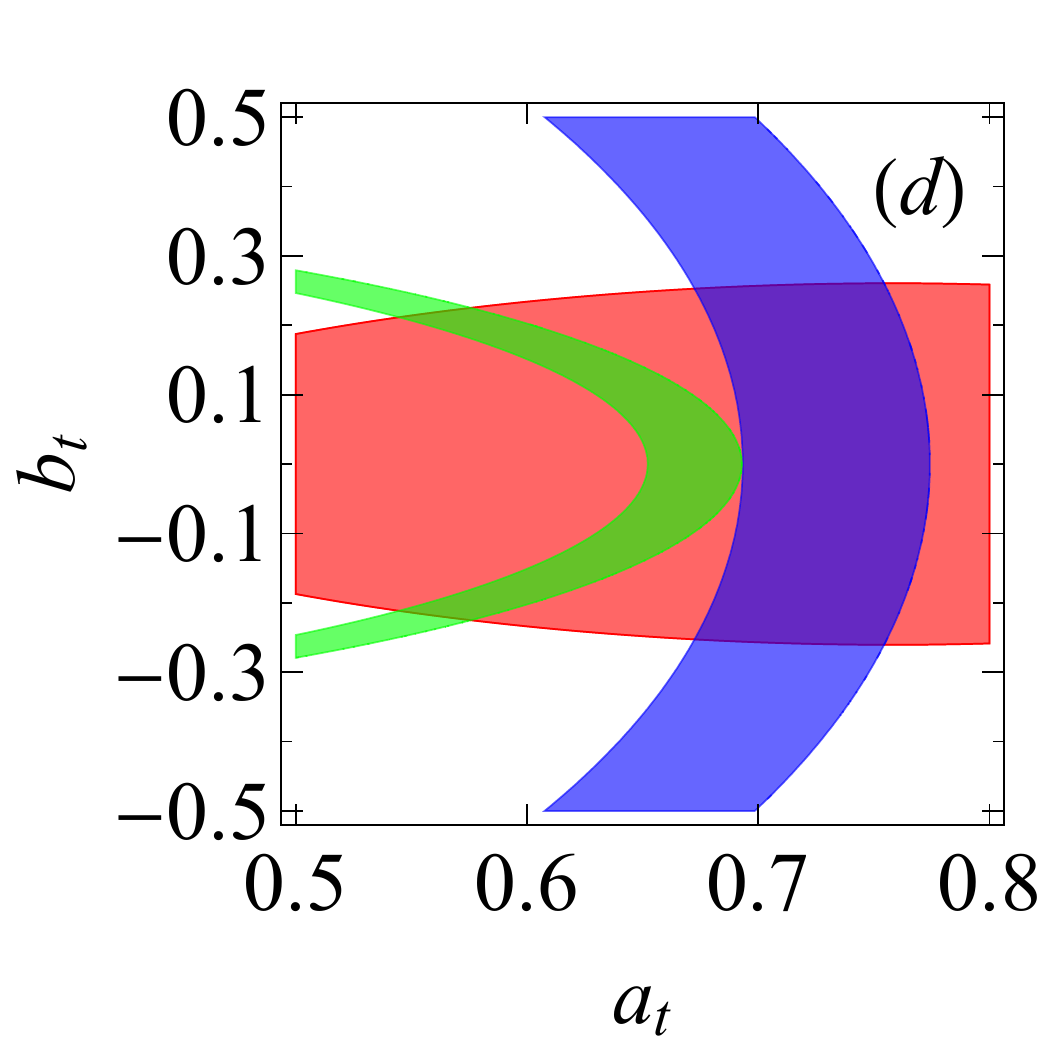}
\caption{Parameter space consistent with the measurement of $ggH(\rightarrow\gamma \gamma)$ (green) and $t\bar{t}H$ (blue) productions at the $1\sigma$ C.L., while the red region is allowed by the $\tttt$ production measurement with $50\%$ uncertainty at the 13 TeV LHC with an integrated luminosity of 3000 fb$^{-1}$. In (a) and (b) the Higgs boson total width is taken to be the same as the SM prediction, while in (c) 1.18 times and (d) 0.54 times as the SM prediction . }
\label{fig:disregion:3000} 
\end{figure*}

The recent measurements of the $t\bar{t}H$ production and the $gg\to H\to\gamma\gamma$ process by the CMS collaborations yields~\cite{CMS:2018dmv,Sirunyan:2018koj}
\begin{align}
& \mu_{t\bar{t}H}^{\rm Exp}=0.96^{+0.34}_{-0.31}, && \mathcal{L}=41.5~{\rm fb}^{-1},\nn\\
& \mu_{gg\to H \to \gamma\gamma}^{\rm Exp}=1.16^{+0.21}_{-0.18}, && \mathcal{L}=36.6~{\rm fb}^{-1}. 
\end{align}
As very complicated statistical methods have been used in the experimental analysis, it is difficult for us to estimate how large the statistical uncertainties should be when collecting more data. Therefore, we adopt the uncertainties shown above to explore the sensitivities of the LHC with an integrated luminosity of $300~{\rm fb}^{-1}$.

We begin with the case of $R_\Gamma=1$, i.e. $\Gamma_H=\Gamma_H^{\rm SM}$. In Fig.~\ref{fig:aliveregion:300}(a) we plot the parameter space in the plane of $a_t$ and $b_t$ that are consistent with the current measurements of the $gg\to H\to\gamma\gamma$ channel (green) and the $t\bar{t}H$ production (blue) at the $1\sigma$ confidence level. The blue eclipse of the $t\bar{t}H$ production centers around $a_t=0$ and $b_t=0$ as expected from Eq.~\ref{eq:tth_th}. On the other hand, the $gg\to H\to \gamma\gamma$ channel is affected by the $y_t$ coupling in both the production and decay. That makes the green eclipse shifts towards to positive $a_t$ region; see Eq.~\ref{eq:ggh_th}. The red region denotes the allowed parameter space at the 95\% C.L. if only null results were reported in the search of the $t\bar{t}t\bar{t}$ event at the 13~TeV LHC with an integrated luminosity of $300~{\rm fb}^{-1}$. Note that the $t\bar{t}t\bar{t}$ production suffers from a large scale dependence in the theoretical prediction of the production cross section. In order to make a conservation estimation of the parameter space, we use the lower value of $\sigma(t\bar{t}t\bar{t})$ allowed by the scale dependence, i.e. $\sigma_0-\delta\sigma_\mu$, in our global analysis. It turns out that the bound from the $t\bar{t}t\bar{t}$ production is loose in comparison with the constraints from the other two channels. As a result, the allowed parameter space is determined by the $gg\to H$ and the $t\bar{t}H$ productions. The bounds on $a_t$ and $b_t$ read as 
\beq
0.73<a_t<1.14,\qquad -0.51<b_t<0.51. 
\eeq
That corresponds to a constraint on the CP-phase of $y_t$ as following:
\beq
-\frac{\pi}{5} \le\theta \le \frac{\pi}{5}.
\eeq

Making use of the fact that the shape of parameter space of the di-photon channel is sensitive to the $y_t$ coupling, we can vary the $R_\Gamma$ parameter to check the consistence of the three channels. As shown in Eq.~\ref{eq:ggh_th}, increasing $\Gamma_H$ requires larger values of $a_t$ and $b_t$ to compensate the suppression effect of $\Gamma_H$ in the denominator. Meanwhile, the cancellation between the $W$-boson loop and  the top-quark loop further distorts the shape of the green eclipse. For a large $R_\Gamma$, one cannot even find a common parameter space consistent with the three channels; see Fig.~\ref{fig:aliveregion:300}(b). That yields a constraint on the Higgs-boson width as 
\beq
R_\Gamma\leq 2.17,
\eeq
or
\beq
\Gamma_H\leq 2.17\times \Gamma_H^{\rm SM}\simeq 8.7~{\rm MeV},
\eeq
which is is comparable to the upper limits of the Higgs total width given by the ATLAS and the CMS collaborations~\cite{Aaboud:2018puo} which determine the Higgs-boson width from the comparison of the on-shell and off-shell effects in the $gg\to H \to ZZ^*$ channel.

Finally, we explore the potential of High luminosity phase of the LHC (HL-LHC) which is going to collect an integrated luminosity of $3000~{\rm fb}^{-1}$. Thanks to the unprecedented high luminosity, the $gg\to H\to \gamma\gamma$ and $t\bar{t}H$ productions can be measured very precisely. For example, it is shown that~\cite{Venditti:2018gre}
\begin{align}
&\mu_{t\bar{t}H}^{\rm HL}=1.00^{+0.11}_{-0.11},\nn\\
&\mu_{gg\to H \to \gamma\gamma}^{\rm HL}=1.00^{+0.05}_{-0.05}.
\end{align}
In addition, the $t\bar{t}t\bar{t}$ production can also be discovered, and we assume the contribution from the Higgs-boson mediation can be measured with an accuracy of $50\%$. 
Figure~\ref{fig:disregion:3000} displays the parameter space in the plane of $a_t$ and $b_t$ allowed by the projected measurements of the $gg\to H\to\gamma\gamma $ process (green), the $t\bar{t}H$ production (blue) and the $t\bar{t}t\bar{t}$ production (red), assuming $R_\Gamma=1$. Now the $t\bar{t}t\bar{t}$ channel yields two isolated ovals in the plane of ($a_t$, $b_t$), of which only the one around the SM value ($a_t=1$ and $b_t=0$) is allowed. See Fig.~\ref{fig:disregion:3000}(b) for the details.
Combing the three channels yields a constraint on $a_t$ and $b_t$ as follows:
\beq
0.93<a_t<1.03,\qquad -0.22<b_t<0.22,
\eeq
which corresponds to 
\beq
-\frac{2\pi}{27} \lesssim\theta \lesssim \frac{2\pi}{27}.
\eeq
Again, setting $R_\Gamma$ as a free parameter sizably alters the parameter space of the diphoton channel and gives rise to a tight constraint on the Higgs-boson width. For $R_\Gamma>1$, the green eclipse stretches toward to large $a_t$ region; see Fig.~\ref{fig:disregion:3000}(c).  A good news is that the HL-LHC is able to probe a small $R_\Gamma$. If the Higgs-boson width is smaller than the SM value, i.e. $R_\Gamma<1$, a smaller values of $a_t$ and $b_t$ are needed and, for that reason, the green eclipse shrinks. See ~\ref{fig:disregion:3000}(d). By varying $R_\Gamma$ we are able to obtain a more stringent bound on $R_\Gamma$ as following:
\beq
0.54 \le R_\Gamma \le 1.18,
\eeq
i.e. 
\beq
2.2~{\rm MeV}\lesssim \Gamma_H\lesssim 4.7~{\rm MeV}. 
\eeq

\noindent\textbf{4. Summary.}

The multi-top production is a powerful tool to probe the CP property of the top-Higgs interaction,  
$$\mathcal{L}_{\htt} = -\frac{m_t}{v}H\bar{t} (a_t+ib_t\gamma_5) t ,$$
and the Higgs-boson width $\Gamma_H$ at the LHC. After considering the large scale uncertainty ($\sim 50\%$) of the signal rate, we showed that the four top-quark production alone can exclude a purely CP-odd top-quark Yukawa coupling ($a_t=0$, $b_t=1$) at the 13~TeV LHC with an integrated luminosity of $230~{\rm fb}^{-1}$. Furthermore, increasing the integrated luminosity to $430~{\rm fb}^{-1}$ excludes a purely CP-odd coupling regardless the size of the Yukawa coupling. 

As the top-quark Yukawa coupling and the Higgs-boson width play different roles in various Higgs-boson productions, combining different channels would impose tighter constraints on the CP-phase $\theta\equiv \arctan( b_t/a_t)$ and Higgs-boson width. In this study we considered three production channels as follows: 
\begin{itemize}
\setlength{\itemsep}{0pt}
\item[(1)] the single-Higgs production with a subsequent decay of $H\to \gamma\gamma$,
\item[(2)] the $t\bar{t}H$ associated production, 
\item[(3)] the four-top production. 
\end{itemize}
After considering the current and projected experimental measurements, we showed that, assuming the Higgs-boson width is not altered by new physics effect, the CP-phase is bounded to be $|\theta| \le \frac{\pi}{5}$, if only null result is found in the search of $t\bar{t}t\bar{t}$ events at the LHC with an integrated luminosity of $300~{\rm fb}^{-1}$. Furthurmore, varying the Higgs-boson width dramatically distorts the parameter space of the $gg\to H\to \gamma\gamma$ channel and even leads to non-consistent parameter space for all the three channels. We found that $\Gamma_H\leq 2.17 \Gamma_H^{\rm SM}\simeq 8.7~{\rm MeV}$. 

At the high luminosity phase of the LHC with an integrated luminosity of $3000~{\rm fb}^{-1}$, all the three channels are expected to be measured precisely. We demonstrated that, if the Higgs-boson related $t\bar{t}t\bar{t}$ production is measured with an uncertainty of $50\%$, the combined analysis yields stringent constraints on the CP-phase as $|\theta|\lesssim 2\pi/27$ with the assumption of $\Gamma_H=\Gamma_H^{\rm SM}$.  Varying the Higgs-boson width yields $2.2~{\rm MeV}\lesssim \Gamma_H\lesssim 4.7~{\rm MeV}$. 

~\\
\noindent{\bf Acknowledgements}

The work is supported in part by the National Science Foundation of China under Grant Nos. 11175069, 11275009, 11422545, 11725520, 11675002, 1163500, 11775093, 11805013 and the Fundamental Research Funds for the Central Universities under Grant No. 2018NTST09.

\bibliographystyle{apsrev}
\bibliography{reference}

\end{document}